\documentclass[12pt]{article}
\usepackage{graphics}
\newcommand{\rlpartial}{\partial^{{}^{{}^{\!\!\!\!\!\!\!\longleftrightarrow}}}}

\begin{document}
\markright{Nonthermal nature of...}
\title{Nonthermal nature of extremal Kerr black holes}
\author{Tony Rothman \thanks{trothman@titan.iwu.edu}\\
~Dept.\ of Physics, Illinois
Wesleyan University, Bloomington, IL 61702, USA\\
and\\
Dept.\ of Maths and Applied Maths,\\
 University of Cape Town, Rondebosch 7700,
Cape, South Africa.\\  }
\date{{\small \LaTeX-ed \today}}
\maketitle

\begin{abstract}
Liberati, Rothman and Sonego have recently showed that objects collapsing
into extremal Reissner-Nordstr\"om black holes do not behave as thermal
objects at any time in their history.  In particular, a temperature, and
hence thermodynamic entropy, are undefined for them.  I demonstrate that the
analysis goes through essentially unchanged for bodies collapsing into
extremal Kerr black holes.

\vspace*{5mm} \noindent PACS: 04.62.+v, 04.70.Dy, 04.20.Dw\\
Keywords: Black holes, Hawking radiation, moving mirrors.
\end{abstract}

\section{Introduction}
\label{sec1}

Recent results in quantum field theory in curved spacetime have demonstrated
fairly conclusively that extremal Reissner-Nordstr\"om (RN) black holes are of a
fundamentally different class than their nonextremal counterparts.  Liberati,
Rothman and Sonego (LRS, \cite{LRS00}) examined spherical bodies collapsing into
extremal Reissner-Nordstr\"om black holes and showed that the object's
particle-emission spectrum is nonthermal. Consequently a temperature is
undefined for them.  Furthermore, at no time during the history of the object
do the properties of the stress-energy tensor of the radiation correspond to
those of a
nonextremal RN hole, and thus extremal holes never behave as thermal
objects.  Simultaneously with the LRS work, Anderson, Hiscock and
Taylor \cite{AHT00} demonstrated that
static extremal RN solutions for ``physically realistic" values of the
curvature coupling constant simply do not exist to perturbative order in
semi-classical gravity.  Taken together, these results and others (see LRS for
references) constitute strong
evidence that the $Q^2 = M^2$ RN solutions, which are the absolute-zero state in
black hole dynamics, are for reasons as yet unclear disallowed by nature.

Of greater astrophysical interest than charged black holes are rotating black
holes.  For that reason it is important to extend the results to the extremal
Kerr metric.  It is well known (see, e.g. Hawking \cite{Hawking75} or DeWitt
\cite{DeWitt75}) that the spectrum of the nonextremal Kerr solution
can be obtained from the Schwarzschild-Droste (SD)\footnote
{Johannes Droste, a pupil of Lorentz, independently
announced the ``Schwarzschild" exterior solution within four months of
Schwarzschild. See, ``The field of a single centre in Einstein's theory of
gravitation, and the motion of a particle in that field," Koninklijke
Nederlandsche Akademie van Wetenschappen, Proceedings {\bf 19}, 197 (1917).}
spectrum merely by taking into
account the frequency shift induced by the rotation of the hole (below).
One suspects that
the extremal Kerr solution can be obtained from the extremal RN solution
in the same way but because previous work has shown that extremality is
dangerous territory, one might want to see this as explicitly as possible.
In this letter I demonstrate that, to be sure, the behavior of the extremal Kerr
solution is essentially identical to that of the RN case.
Thus it appears that the nonthermal nature of extremal black holes is a
generic property.  In retrospect, this is to have been
expected because the nature of the horizon
changes completely at extremality \cite{Boyer69}.

LRS used the well known ``moving mirror" technique to model the collapse.
Rather than deal with the full four-dimensional problem one considers a
one-dimensional mirror moving in two-dimensional Minkowski space.
Massless scalar fields are propagated inwards from ${\cal I^-}$
and bounced off the mirror
to ${\cal I^+}$.  Due to the mirror's
motion, one expects that the In and Out vacuum states will differ, leading to
particle production.  The spectrum is a function of the mirror's trajectory,
and so a key step in the calculation is to determine the worldline of the
star's center just before formation of the horizon.  This late-time trajectory
is taken in turn to correspond to the asymptotic trajectory of the mirror.
Unfortunately, Kruskal corrdinates, which are typically employed in such
calculations and which contain the surface gravity $\kappa$, break down in
the extremal limit where $\kappa$ vanishes (below).
LRS were able to provide a simple coordinate extension for this case, which
allowed them to calculate the late-time history of the center of the collapsing
object.  Remarkably, this turned out to correspond precisely
to that of a uniformly accelerated mirror in Minkowski space.  The properties
of such a system are well known \cite{Birrell82}.  In particular, the emission
spectrum is nonplanckian,
demonstrating that temperature is undefined for these objects.  I now show
that the same applies to the Kerr solution.

\section{Coordinate extension for the extremal
Kerr solution} \setcounter{equation}{0} \label{sec2}

The Kerr metric in Kerr coordinates is
\begin{eqnarray}
ds^2 &=&  -[1-2Mr\rho^{-2}]dv^2 + 2drdv+\rho^2d\theta^2
    +\rho^{-2}[(r^2+a^2)^2 \nonumber\\
    &&- \Delta a^2 sin^2\theta]sin^2\theta d\phi_*^2
    -2a sin^2\theta d\phi_*dr - 4Mra\rho^{-2}sin^2 \theta d\phi_* dv,
    \label{Kerr}
\end{eqnarray}
where as usual
\begin{equation}
dv = dt + \frac{(r^2 + a^2)dr}{\Delta} \ ; \
                d\phi_* = d\phi + \frac{a dr}{\Delta}
\label{dv}
\end{equation}
and
\begin{equation}
\rho^2 = r^2 + a^2cos\theta \ ; \ \Delta = r^2 -2Mr +a^2 \ .
\end{equation}
Note that $dv$ represents an advanced null coordinate corresponding to the
$dv = dt + dr_*$ employed in SD or RN calculations, and that
$a$ is the angular momentum parameter.

To find the late-time history of the collapsing body's center, it is useful to
employ double-null coordinates and so we define
\begin{equation}
du = dt - \frac{(r^2 + a^2)dr}{\Delta}
\end{equation}

Before rewriting the Kerr metric in terms of $u,v$, it is important to verify
that these coordinates are continuous in the extremal limit $a=M$.  This is
easily done.  Integrating the expression for $dv$ in (\ref{dv}) yields
for the nonextremal case:
\begin{equation}
\begin{array}{lll}
v(a,M) &\equiv& t + r_*(a,M)\vspace{2 mm}\nonumber\\
    &=& t + \frac{1}{2\sqrt{M^2 -a^2}}\{(r_+ - r_-)r
        +(r_+^2 +a^2)\ln (r-r_+) \vspace{2 mm}\nonumber\\
        &&-(r_-^2 + a^2)\ln(r-r_-)\} + {\rm const}\ .
\label{vaM}
\end{array}
\end{equation}
Here, $r_\pm \equiv M \pm \sqrt{M^2 -a^2}$.

The extremal value for $v$ is found by setting $a=M$ in the expression for $dv$
{\em before} integrating.  Then $\Delta = (r-M)^2$ and
\begin{equation}
dv(M,M) = dt + \frac{(r^2 +M^2)dr}{(r-M)^2}.
\end{equation}
Integrating gives
\begin{eqnarray}
v(M,M) &\equiv& t +r_*(M,M)\nonumber\\
    &=& t + r + 2M\ln (r-M) - \frac{2M^2}{(r-M)} + {\rm const}.
\label{vMM}
\end{eqnarray}
Note that (\ref{vaM}) appears to give $0/0$ when $a=M$, whereas (\ref{vMM}) is
well behaved except at $r=M$, the extremal horizon.  Nevertheless, if one
sets $M = a + \epsilon, \ \epsilon << a$ and works to first order in $\epsilon$,
it is straightforward to show that (\ref{vMM}) is the limit of (\ref{vaM}) as
$\epsilon \rightarrow 0$.  Therefore, the coordinate $v$ is continous
at extremality.  The same holds for $u$ and $\phi_*$.

Since $u$ and $v$ are good at extremality, we now specialize to the case $a=M$.
The metric (\ref{Kerr}) becomes
\begin{eqnarray}
ds^2 &=& -[1 - \frac{2Mr}{\rho^2}]dv^2 + 2dr dv + \rho^2 d\theta^2
    +\frac{1}{\rho^2}[(r^2 + M^2)^2 - (r-M)^2M^2sin^2\theta]
        sin^2\theta d\phi_*^2 \nonumber\\
    &&-2Msin^2\theta dr d\phi_* -\frac{4M^2r}{\rho}\sin^2\theta d\phi_* dv
    \ .
\end{eqnarray}
Eliminating $dr$ in favor of $du$ and $dv$ yields
\begin{equation}
\begin{array}{lll}
ds^2 &=& [\frac{(r-M)^2}{(r^2 + M^2)} + (\frac{2M}{r}-1)]dv^2
    - \frac{(r-M)^2}{(r^2 + M^2)} du dv \vspace {2 mm} \nonumber\\
      &&+ \frac{1}{\rho^2}[(r^2+M^2)^2-(r-M)^2M^2sin^2\theta]sin^2\theta
            d\phi_*^2 \vspace {2 mm}\nonumber\\
     && -[\frac{4M^2r}{\rho^2} + \frac{M(r-M)^2}{(r^2 + M^2)}]
        sin^2\theta d\phi_*dv
      + \frac{M(r-M)^2}{(r^2 + M^2)}sin^2\theta
        d\phi_* du +\rho^2 d\theta^2 \ .
      \label{dsM}
\end{array}
\end{equation}

This metric is evidently degenerate at $r = M$
and so the coordinates $u,v$
are not good on the horizon at extremality.  However, the
$(r-M)^2$ degeneracy is of
exactly the same form LRS found for the extremal RN case.  Thus, the same
coordinate extension should work here as well.  We define a transformation
\begin{equation}
\psi(\xi)= 4M\left(\ln\xi - \frac{M}{2\xi}\right) \label{psi}
\end{equation}
and assume
\begin{equation}
\begin{array}{lll}
u = -\psi(-{\cal U}) &\Leftrightarrow &{\cal U} = -\psi^{-1}(u)\\
v = \psi({\cal V}) &\Leftrightarrow &{\cal V} = \psi^{-1}(u)
\end{array}
\label{Gen}
\end{equation}
where \ ${\cal U}$ and ${\cal V}$ are the new coordinates that should be
regular on the horizon at extremality.

To motivate this choice of $\psi$,
recall that $v,u$ are constructed adding or subtracting $r_*$ to $t$.
In the SD case, $r_*$ has only a logarithmic divergence at $r = 2M$,
as can be seen by setting $a=0$ in (\ref{vaM}). (The same holds for
nonextremal RN). If one took only the first term in (\ref{psi}),
$\psi(u) = 4M\ln({\cal U})$, then one would have ${\cal U} = -e^{-\kappa u}
\ ;\ \kappa = 1/4M$,
which is exactly the Kruskal transformation. However, for charged, rotating
holes
$\kappa \sim (M^2 - a^2 - Q^2)$.  Thus, in the extremal limit,
the surface gravity vanishes and, as mentioned above, the Kruskal
transformation becomes ill-defined.
From (\ref{vMM}), however, we observe that at extremality,
$r_*$ has not only the logarithmic divergence of the nonextremal case
but the added pole $(2M^2)/(r-M)$.  We therefore pick the simplest
possible generalization
of the Kruskal transformation and add this term to $\psi$ to get (\ref{psi}).
Note also that $\psi'(\xi)=4M/\xi+2M^2/\xi^2
> 0$, always, and so $\psi$ is monotonic; therefore
(\ref{Gen}) is a well-defined coordinate transformation. From (\ref{vMM})
we have at extremality
\begin{equation}
r_*(M,M) = r+\frac{1}{2}\psi (r-M) ,
\end{equation}
Near the horizon $v$ is constant, which implies $t \sim -r_*$ and therefore
$u \sim -2r* \sim -\psi(r-M)$.  Thus, ${\cal U} =
-\psi^{-1} \sim -(r-M)$ and the derivative
$\psi'(-{\cal U}) \sim 4M^2/(r-M)^2$. It is this factor of $(r-M)^2$ in the
denominator that will kill the $(r-M)^2$ factors in the numerator of (\ref{dsM}),
for when we rewrite the metric in terms of the new variables ${\cal U, V}$ we
find
\begin{equation}
\begin{array}{ll}
ds^2 \approx & [\frac{(r-M)^2}{(r^2+M^2)}
        + (\frac{2M}{r} -1)]\psi'({\cal V})^2 d{\cal V}^2
    - \frac{4M^2}{(r^2 +M^2)} \psi'({\cal V})d{\cal U}d{\cal V}
        \vspace{3 mm}\nonumber\\
    & +\frac{1}{\rho^2}[(r^2+M^2)2 - (r-M)^2M^2sin^2\theta]
        sin^2\theta d\phi_*^2 \vspace{3 mm}\nonumber\\
    &-[\frac{4M^2r}{\rho} + \frac{M(r-M)^2}{(r^2 + M^2)}]sin^2\theta
    d\phi_*\psi'({\cal V})d{\cal V}
    +\frac{4M^3}{(r^2+M^2)}d\phi_* d{\cal U} + \rho^2d\theta^2
\end{array}
\end{equation}
Note that since ${\cal V}$ is everywhere nonzero and finite, $\psi'({\cal V})$
is regular on the horizon.  Thus nothing particular happens at $r=M$ and
$\psi$ is a good coordinate transformation there.
Note also that the degeneracy on the horizon is due solely to the behavior of
the coordinate $u$.  For this reason the coordinate extension $\psi$
is good for any value of $\theta$ and the results are
completely general.

\section{Late-time history}
\setcounter{equation}{0}
 \label{sec3}
We now use $\psi$ to construct the late-time history of the collapsing
star's center in the coordinates $u,v$. LRS performed such a calculation
 for the case of
spherical symmetry.  Of course the Kerr metric is only axially symmetric,
not spherically symmetric,
but if we specialize to the
equatorial plane $\theta = \pi/2$, then the problem becomes two-dimensional
and can be treated in
the same way.  Moreover, as long as we are concerned solely with the
{\em center} of the star (as opposed to, say, the surface), all angular
coordinates become degenerate and cannot
be relevant. Thus, even for arbitrary $\theta$, one evidently needs
to consider only $r$ and $t$ (or $u$ and $v$), as LRS did
for the RN case.

The coordinates $u$ and $v$ are valid only outside the collapsing star and so
the trajectory of the center, we must extend them to the interior.
This is accomplished in standard
fashion \cite{Birrell82}.  Ignoring the angular variables,
the interior can be described by some interior null coordinates, $U$ and $V$
and the metric components $g_{\mu\nu}(U,V)$ can be chosen to be regular on
the horizon. The center of the star can be taken at radial coordinate = 0,
in which case
$V = U$ and ${\rm d}V = {\rm d}U$ there.

Because the Kruskal-like coordinates $\cal U$ and $\cal V$ are regular
everywhere as well, they can be matched to $U$ and $V$. In
particular, if two nearby outgoing rays differ by ${\rm d}U$
inside the star, then they will also differ by ${\rm
d}U=\beta({\cal U}) {\rm d}{\cal U}$, with $\beta$ a regular
function, outside. By the same token, since $V$ and $v$ are
regular everywhere, we have ${\rm d}V = \zeta(v) {\rm d}v$, where
$\zeta$ is another regular function. In fact, if we consider the
last ray $v=\bar v$ that passes through the center of the star
before the formation of the horizon, then to first order ${\rm d}V
= \zeta (\bar v){\rm d}v$, where $\zeta (\bar v)$ is now constant.

We can write near the horizon
\begin{equation}
{\rm d}U = \beta(0)\frac{{\rm d}{\cal U}}{{\rm d}u}{\rm d}u\;.
\end{equation}
Since for the center of the star ${\rm d}U = {\rm d}V = \zeta(\bar
v){\rm d}v$, this immediately integrates to
\begin{equation}
\zeta(\bar v)(v - {\bar v}) = \beta (0){\cal U}(u) =
         -\beta (0) \psi^{-1}\left(-u\right)
        \sim -2 \beta (0)\frac{M^2}{u}\;.
    \label{zeta}
\end{equation}
The last approximation follows from Eq.\ (\ref{psi}) where
$\xi\sim \psi^{-1}(-2M^2/\xi)$ near the horizon.

Thus the late-time worldline for the center of the star is,
finally, represented by the equation
\begin{equation}
v \sim {\bar v} - \frac{A}{u}\;,\quad u\to +\infty\;, \label{traj}
\end{equation}
where $A = 2\beta (0)M^2/\zeta(\bar v)$ is a positive constant
that depends on the details of the internal metric and
consequently on the dynamics of collapse.\footnote {One might
wonder why we did not consider $dV = \zeta({\cal V})d{\cal V}$
instead of ${\rm d}V = \zeta(v) {\rm d}v$.  In the former case we
do not know the function $\zeta(\cal V)$ and so the above
integration could not be carried out.} This worldine is obviously
hyperbolic and corresponds identically to that of a uniformly
accelerated mirror in Minkowski space.  This should be contrasted
with the late-time worldline of a nonextremal hole, as can be
found by using the ``Kruskal-part" of the transformation $\psi$ in
(\ref{zeta}).
\begin{equation}
v\sim {\bar v}-B{\rm e}^{-\kappa u}\;,\quad u\to +\infty\;.
\label{nextraj}
\end{equation}

As shown by LRS for the case of RN holes,
the hyperbolic worldline cannot be obtained as an
approximation of the nonextremal worldline (\ref{nextraj}) and so
(\ref{traj}) represents a true discontinuity between extremal RN holes
and their nonextremal counterparts.  We now see the same conclusion applies to
extremal Kerr holes.

With the benefit of hindsight this result is perhaps not so surprising, at least
for the RN case.
Consider the typical homework problem of
dropping a rock into a black hole from some $r = R_o$.  One generally solves
this by starting from the condition on the four-velocity $u_\mu u^\mu = - 1$
 and the condition
$u_o = -\tilde E$ = energy per unit mass = constant.  For RN one
has $u_o = -(1-2M/r + Q^2/r^2)u^o$ and $u_r = (1-2m/R
+Q^2/r^2)^{-1}u^r$, with $u^o = dt/d\tau$ and $u^r = dr/d\tau$.
Forming $dr/dt$ for the extremal RN metric, one gets after a few
elementary substitutions
\begin{equation}
dt = \frac{M z^3 dz}{(z-1)^3[z^2 - e^2]^{1/2}}
\end{equation}
where $e\equiv \tilde E^{-1}$ and $z \equiv 1 + M/(r-M)$.  The integral is
cumbersome but can be evaluated analytically in terms of elementary functions.
The result near the horizon ($z >> 1$) is
\begin{equation}
t = k_1 + Mz + 2M\ln z +\frac{k_2}{z} + {\cal O} (\frac{1}{z^2})
\end{equation}
where $k_1$ and $k_2$ are constants.  For the extremal RN
solution, $r_* = r - M^2/(r-M) + 2M\ln (r-M)$.  Near the horizon
($r=M$) the $M^2$ term dominates.  Using this and the fact that in
the same regime $t \approx -r_* \Rightarrow u \approx -2r_*$, one
brings the above equation into the form
\begin{equation}
v \approx k_1 + \frac{k_3}{u},
\end{equation}
again a hyperbolic worldline, as distinct from the exponential form similar
to (\ref{nextraj}) that one finds
for a rock falling into a SD or nonextremal RN black hole.
 As mentioned in the Introduction, the horizon structure changes completely
 at extremality.  Moreover, the horizon itself is moved
in to $r=M$, and the shape of the effective potential is totally altered.
Given all this, it should really come as no suprise that the trajectories of
objects
falling into extremal versus nonextremal holes should differ.

\section{Nonthermal Spectrum}
\setcounter{equation}{0} \label{sec4}

To implement the moving mirror analogy, one requires that the wave
equation $\phi_{;\mu}\;^{;\mu} = 0$ be separable, which implies
that $\phi = f(u) + g(v)$. Then one move the problem to Minkowski
space, where the null coordinates are now assumed to be $u = t -
x$ and $v = t + x$.  The mirror trajectory is taken to be that of
the center of the star, $v = p(u)$ .  With the boundary conditions
that one has pure ingoing solutions on ${\cal I^-}$ and pure
outgoing solutions on ${\cal I^+}$ and that there be total
reflection at the mirror, i.e., $\phi(p(u),u) = 0$, it is easy to
show that
\begin{equation}
\phi_\omega^{\rm (in)}(u,v)={{\rm i}\over\sqrt{4\pi\omega}}
\left({\rm e}^{-{\rm i}\omega v}-{\rm e}^{-{\rm i}\omega
p(u)}\right) \label{in}
\end{equation}
and
\begin{equation}
\phi_\omega^{\rm (out)}(u,v)={{\rm i}\over\sqrt{4\pi\omega}}
\left({\rm e}^{-{\rm i}\omega u}- \Theta\left(\bar{v}-v\right){\rm
e}^{-{\rm i}\omega q(v)}\right)\;. \label{out}
\end{equation}
Here $q(v)=p^{-1}(v)$ and $\omega>0$.  (See LRS for further details.)

The relevant Bogoliubov coefficient is defined as
\begin{equation}
\beta_{\omega\omega'}=-\left(\phi_\omega^{\rm (out)},
\phi_{\omega'}^{\rm (in)\ast}\right)= {\rm i}\int_0^{+\infty}{\rm
d}x\left[\phi_\omega^{\rm
(out)}(u,v)\rlpartial_t\phi_{\omega'}^{\rm
(in)}(u,v)\right]_{t=0}\;. \label{betaa}
\end{equation}
which gives a particle spectrum
\begin{equation}
\langle N_\omega\rangle=\int_0^{+\infty}{\rm
d}\omega'\,|\beta_{\omega\omega'}|^2\;. \label{spectrum}
\end{equation}

For SD and nonextremal RN, it is the evaluation of (\ref{betaa})
and (\ref{spectrum}) that yields the black-body spectrum because the
function $p$ is exponential at late times.  For the extremal RN case, however,
one uses the hyperbolic form, equivalent to a uniformly accelerated mirror.
The properties of the uniformly accelerated mirror have been extensively
 studied and the Bogoliubov coefficients have been calculated \cite{Birrell82}.
 In the
asymptotic regime for the trajectory (\ref{traj})
\begin{equation}
\beta_{\omega\omega'}\approx \frac{\sqrt A}{\pi}{\rm e}^{{\rm
i}{\sqrt A}(\omega-\omega')}
                K_1(2(A\omega\omega')^{1/2})\;,
\label{beta}
\end{equation}
where $K_1$ is a modified Bessel function. For
argument $z$, $K_1(z)\sim 1/z$ for $z\to 0$, and
$K_1(z)\sim\sqrt{\pi/(2z)}\,{\rm e}^{-z}$ when $z\to +\infty$
\cite{as}

 For the extremal
Kerr black hole the analysis goes through almost unchanged.
The solutions to wave equation for Kerr can in fact be separated
into the above form \cite{PB}, where now
$u$ and $v$ are exactly the null coordinates we have been using.  In the
vicinity of the Kerr horizon, the radial wave function has the form
$R \sim exp (\pm i\tilde\omega r_*)$.  Here, $\tilde\omega \equiv
\omega - m\Omega$, $m$ is the
azimuthal quantum number and $\Omega = 1/(2M)$ is the angular velocity of the
 black hole on the horizon at extremality.  The outgoing modes
 $\phi^{(out)} \sim exp(-i \tilde\omega u)$ get transported back to
 to become $\phi^{(out)} \sim exp (i \tilde\omega q(v))$ on ${\cal I^-}$.
   The Bogoliubov coefficient
 is essentially the Fourier transform of this quantity,
 \begin{equation}
 \beta_{\tilde\omega'\tilde\omega}
     \approx \int exp\{i\tilde\omega'v + i\tilde\omega q(v)\} dv
\end{equation}
which is the relevant integral one encounters while evaluating (\ref{betaa}).
Thus, since we have established that $p(u)$ and hence $q(v)$ are
the same for both extremal Kerr and extremal RN, the Bogoliubov coefficients
are also the same with the replacement $\omega \rightarrow \tilde\omega$.
This gives the frequency shift mentioned in the Introduction.
Consequently, with the substitution $\tilde\omega$ for $\omega$, the extremal
Kerr spectrum is found to be identical to the extremal RN spectrum.

 The main point is that since the spectrum goes
 like the square of a $K_1$ Bessel
 function it is definitely not a black-body spectrum.  Therefore temperature is
 simply undefined.  Moreover, as shown by LRS the variance of the stress-energy
 tensor decays to zero as the power law
\begin{equation}
\langle :\!T^2_{uu}\!:\rangle=-{A^2\over 4\pi^2
\left(A+\left(v-\bar{v}\right)u\right)^4}\sim - {A^2\over
4\pi^2\left(v-\bar{v}\right)^4u^4}\;. \label{powlaw}
\end{equation}
as distinct from the nonextremal black hole, the variance of whose
stress-energy tensor decays exponentially:
\begin{equation}
\langle :\!T_{uu}^2\!:\rangle={1\over (4\pi)^2}\left(
{\kappa^4\over 48}- {4\kappa^2B^2{\rm e}^{-2\kappa u}\over\left(
v-\bar{v}+B{\rm e}^{-\kappa u}\right)^4}\right) \sim
{\kappa^4\over 768\pi^2}- {\kappa^2B^2{\rm e}^{-2\kappa u}\over
4\pi^2\left(v-\bar{v}\right)^4}\;. \label{explaw}
\end{equation}
Contrary to the extremal case, the ``nonextremal" variance $\Delta
T_{uu}$ tends not to zero as $u\to +\infty$, but to the value
$\kappa^2\sqrt{2}/(48\pi)$, which corresponds to thermal emission.

\section{Conclusions}
\setcounter{equation}{0} \label{sec7}

All these results, including those of \cite{AHT00}, demonstrate that at no time
in their history do extremal black holes act
as thermal objects .  The conclusion would seem
to have significant implications in regard to string theory.  From the point of
 view of thermodynamics,
if temperature is undefined, entropy is also undefined.
Yet string caclulations \cite{Strom96},
which involve only state-counting arguments, indicate that extremal black hole
solutions possess the usual Bekenstein-Hawking entropy. The results of
semi-classical gravity and string theory would appear, then, to be in direct
contradiction.  The resolution to this dilemma is far from clear.  The
calculations discussed in this letter assume $a = M$ configurations
exist, but ignore back-reaction to the metric (or
recoil of the mirror).  Evidently, extraordinary fine-tuning early on
is required to produce extremal solutions
and it may well be that, taking such backreaction into account, the formation of
extremal objects is simply disallowed by nature.  If true, string-theory
calculations refer to impossible objects.  On the other hand, if one cannot
ignore back-reaction---at least at late times---then all semi-classical
calculations
to date (including Hawking's) are incorrect. Presently, about
all that can be said is that extremal black holes seem to represent the limits
 of our
current theories of quantum gravity.

\noindent{\bf Acknowledgements}
It is a pleasure to thank George Ellis and Jeff Murugan for stimulating
conversations and helpful comments on the manuscript.

\end{document}